\documentclass[aps,prb,twocolumn,groupedaddress,showpacs]{revtex4}

\usepackage{graphicx}
\usepackage{dcolumn}
\usepackage{bm}
\def\xsum{\mathop{\sum\nolimits'}}

\begin{document}

\title{On the BCS-BEC crossover in the 2D Asymmetric Attractive Hubbard Model}

\author{Agnieszka Kujawa-Cichy}
\email{agnieszkakujawa2311@gmail.com}
\affiliation{Solid State Theory Division, Faculty of Physics, Adam Mickiewicz University, Umultowska 85,
61-614 Pozna\'n, Poland}
\begin{abstract}
We analyze the evolution from the weak coupling (BCS-like limit) to the strong coupling limit of tightly bound local pairs (LP's) in the 2D asymmetric attractive Hubbard model, in the presence of the Zeeman magnetic field ($h$). The broken symmetry Hartree approximation is used. We also apply the Kosterlitz-Thouless (KT) scenario to determine the phase coherence temperatures. We obtain that for the spin dependent hopping integrals ($t^{\uparrow}\neq t^{\downarrow}$) the homogeneous polarized superfluid (SC$_M$) phase in the ground state for the strong attraction and lower filling can be stabilized. We find a topological quantum phase transition (Lifshitz type) from the unpolarized superfluid phase (SC$_0$) to SC$_M$ and tricritical point in the ($h-\mu$) and spin polarization ($P$) vs. attraction ($U<0$) ground state phase diagrams. The finite temperatures phase diagrams for $t^{\uparrow}\neq t^{\downarrow}$ are constructed. 
\end{abstract}

\maketitle

\section{Introduction}

Unconventional superconductivity in strongly correlated electron systems and spin-polarized superfluidity (in the context of cold atomic Fermi gases) are currently investigated and widely discussed in the leading world literature. 

 Recent works of experimental groups from MIT \cite{ketterle, ketterle2} and also from Rice University \cite{li} began investigations of quantum Fermi gases ($^{6}$Li) with unequal numbers of fermions with down ($\downarrow$) and up ($\uparrow$) spins ($N_{\downarrow}\neq N_{\uparrow}$ -- systems with \emph{population imbalance}). The possibility to control the population imbalance and the coupling has motivated the attempts to understand the BCS-BEC crossover phase diagrams at zero and finite temperatures for imbalanced systems.

The presence of a magnetic field, population imbalance or mass imbalance introduces a mismatch between the Fermi surfaces (FS). This makes the realization of many interesting phases possible, e.g.: the spatially homogeneous spin-polarized superconductivity (breached pair (BP)) which has a gapless spectrum for the majority spin species \cite{Sheehy}. The coexistence of the superfluid and the normal component in the isotropic state is characteristic for the BP phase. This kind of state was originally considered by Sarma \cite{Sarma}. 

Some theoretical studies of Fermi condensates in systems with spin and mass imbalances have shown that the BP state can have excess fermions with two FS's (BP-2 or interior gap state) \cite{Wilczek, Wilczek2, Iskin, Iskin2}. According to some investigations, the interior gap state \cite{Wilczek} is unstable even for large mass ratio \cite{Parish, parish2}. Therefore, the problem of stability of the BP-2 state is open. At strong attraction, the $SC_M$ phase occurs in three-dimensional imbalanced Fermi gases \cite{Sheehy}, \cite{Parish} as well as in the spin-polarized attractive Hubbard model in a dilute limit (for $h\neq 0$, $r=1$ \cite{kujawa2} and $r\neq 1$ \cite{kujawa3}). This homogeneous magnetized superfluid state consisting of a coherent mixture of LP's (hard-core bosons)  and excess spin-up fermions (Bose-Fermi mixture) can only have one Fermi surface (BP-1).

In this paper we focus on $s$-wave superconducting (SC) phases on a square lattice, described by the attractive Hubbard model (AHM) ($U<0$) in a magnetic field with spin dependent hopping \cite{ACichy}:
\begin{equation}
\label{ham}
H=\sum_{ij\sigma} (t_{ij}^{\sigma}-\mu \delta_{ij})c_{i\sigma}^{\dag}c_{j\sigma}+U\sum_{i} n_{i\uparrow}n_{i\downarrow}-h\sum_{i}(n_{i\uparrow}-n_{i\downarrow}),
\end{equation}
where: $\sigma=\uparrow,\downarrow$, $n_{i\uparrow}=c_{i\uparrow}^{\dag}c_{i\uparrow}$, $n_{i\downarrow}=c_{i\downarrow}^{\dag}c_{i\downarrow}$, $t_{ij}^{\sigma}$ -- spin dependent hopping integrals, $U$ -- on-site interaction, $\mu$ -- chemical potential. The Zeeman term can be created by an external magnetic field (in ($g \mu_B \slash 2$) units) or by a spin population imbalance in the context of cold atomic Fermi gases.

Applying the broken symmetry Hartree approximation, we obtain the equations for the gap parameter $\Delta=-\frac{U}{N}\sum_i \langle c_{i \downarrow} c_{i \uparrow} \rangle$, particle number $n=n_{\uparrow}+n_{\downarrow}$ (determining $\mu$), where: $n_{\sigma}=1/N\sum_{i}\langle c^{\dagger}_{i\sigma}c_{i\sigma}\rangle$ and magnetization $M$ \cite{Kujawa, Kujawa5}. The equations take into account the spin polarization ($P=(n_{\uparrow}-n_{\downarrow})/n$) in the presence of a magnetic field and spin-dependent hopping ($t^{\uparrow}\neq t^{\downarrow}$, $\frac{t^{\uparrow}+t^{\downarrow}}{2}=t$, $t^{\uparrow}/t^{\downarrow}\equiv r$).

We also calculate the superfluid stiffness $\rho_s(T)$ which for $t^{\uparrow} \neq t^{\downarrow}$ takes the form:
\begin{widetext}
\begin{eqnarray}
\label{ro_s}
\rho_s(T)&=&\frac{1}{4N}\sum_{\vec{k}} \Bigg\{ \frac{\partial ^{2}\epsilon^{+}_{\vec{k}}}{\partial k_x^2} -\frac{1}{2}\Bigg[\frac{\partial ^{2}\epsilon^{-}_{\vec{k}}}{\partial k_x^2}+\frac{\epsilon^{+}_{\vec{k}}}{\omega_{\vec{k}}}   \Bigg(\frac{\partial ^{2}\epsilon^{+}_{\vec{k}}}{\partial k_x^2}\Bigg) 
+\Bigg(\frac{\partial \epsilon^{-}_{\vec{k}}}{\partial k_x}\Bigg)^2 \frac{|\Delta|^2}{\omega_{\vec{k}}^3} \Bigg] \tanh \Bigg(\frac{\beta E_{\vec{k}\uparrow}}{2}\Bigg)\nonumber \\
&+&\frac{1}{2}\Bigg[\frac{\partial ^{2}\epsilon^{-}_{\vec{k}}}{\partial k_x^2}-\frac{\epsilon^{+}_{\vec{k}}}{\omega_{\vec{k}}} \Bigg(\frac{\partial ^{2}\epsilon^{+}_{\vec{k}}}{\partial k_x^2}\Bigg)  
-\Bigg(\frac{\partial \epsilon^{-}_{\vec{k}}}{\partial k_x}\Bigg)^2 \frac{|\Delta|^2}{\omega_{\vec{k}}^3} \Bigg] \tanh \Bigg(\frac{\beta E_{\vec{k}\downarrow}}{2}\Bigg)\nonumber \\
&+&\Bigg[\frac{\partial \epsilon^{+}_{\vec{k}}}{\partial k_x}+ \frac{\epsilon^{+}_{\vec{k}}}{\omega_{\vec{k}}} \Bigg(\frac{\partial \epsilon^{-}_{\vec{k}}}{\partial k_x}\Bigg)\Bigg]^2 
\frac{\partial f(E_{\vec{k}\uparrow})}{\partial E_{\vec{k}\uparrow}}+\Bigg[\frac{\partial \epsilon^{+}_{\vec{k}}}{\partial k_x}- \frac{\epsilon^{+}_{\vec{k}}}{\omega_{\vec{k}}} \Bigg(\frac{\partial \epsilon^{-}_{\vec{k}}}{\partial k_x}\Bigg)\Bigg]^2 \frac{\partial f(E_{\vec{k}\downarrow})}{\partial E_{\vec{k}\downarrow}}  \Bigg\},
\end{eqnarray}
\end{widetext}
where: $\epsilon_{\vec{k}}^{+}=\frac{\xi_{\vec{k}\uparrow}+\xi_{\vec{k} \downarrow}}{2}$, $\epsilon_{\vec{k}}^{-}=\frac{\xi_{\vec{k}\uparrow}-\xi_{\vec{k} \downarrow}}{2}-\bar{h}$, $\xi_{\vec{k} \sigma}=\epsilon_{\vec{k} \sigma} -\bar{\mu}$, $\epsilon_{\vec{k} \sigma}=-2t^{\sigma}\Theta_{\vec{k}}$, $\Theta_{\vec{k}}=\sum_{l=1}^{d} \cos(k_l a_l)$ ($d=2$ for two-dimensional lattice), $a_l=1$ in further considerations, $\bar{\mu}=\mu-\frac{Un}{2}$, $n=n_{\uparrow}+n_{\downarrow}$ -- particle number, $\bar{h}=h+\frac{UM}{2}$, $M=n_{\uparrow}-n_{\downarrow}$ -- spin magnetization, $E_{\vec{k}\uparrow, \downarrow}= \pm
\epsilon_{\vec{k}}^{-}+\omega_{\vec{k}}$, $\omega_{\vec{k}}=\sqrt{(\epsilon_{\vec{k}}^{+})^2+|\Delta|^2}$, $\beta=1/k_B T$.

For $d=2$, $h=0$, $r=1$, the transition from the superconducting (SC) to the normal (NO) state in the AHM is of the KT type if $n\neq 1$. The KT temperature ($T_c^{KT}$) can be determined from the universal relation:
\begin{equation}
\label{KT}
 k_B T_c^{KT}=\frac{\pi}{2} \rho_s (T_c^{KT}).
\end{equation}

In the strong coupling limit ($|U| \gg t$), AHM ($U<0$, $h=0$, $r\neq 1$) is mapped (via the canonical transformation \cite{Robak, Cazalilla}) onto the pseudo-spin model (with the Hamiltonian operating in the subspace of states without single occupancy). After the transformation to the bosonic operators, this Hamiltonian describes a system of hard-core bosons on a lattice \cite{ACichy, przypis1}: 
\begin{equation}
\label{pseudospin}
H=-\frac{1}{2}\xsum_{i,j}J_{ij} (b_i^{\dag}b_j +h.c.) +\xsum_{i,j} K_{ij}n_i n_j
-\tilde{\mu} \sum_i n_i,
\end{equation}
with the commutation relations \cite{Micnas2,Micnas3,MicnasModern}: $[b_i,b_j^{\dag}]=(1-2n_i)\delta_{ij}$, $b_i^{\dag}b_i + b_ib_i^{\dag}=1$, where $n_i=b_i^{\dag}b_i$. $J_{ij}=2\frac{t_{ij}^{\uparrow} t_{ij}^{\downarrow}}{|U|}$,  $K_{ij}=2\frac{(t_{ij}^\uparrow)^2+(t_{ij}^{\downarrow})^2}{2|U|}$, $\tilde\mu=2\mu+|U|+K_0$ -- chemical potential for bosons, $K_0=\sum_{j} K_{ij}$, primed sum excludes terms with $i=j$.
If $r\neq 1$, the charge density wave ordered (CO) state can develop for any particle concentration. The SC to CO transition is a first order at $h=0$, $r\neq 1$ and $n \neq 1$. The critical $n$ ($n_c$) (within the mean field (MF) approximation) above which SC can coexist with commensurate CO is given by \cite{MicnasModern, Dao}: $n_c=1\pm \Big| \frac{r-1}{r+1} \Big|$.
 
\section{Numerical Results}
Here we continue our analysis performed in Ref. \cite{ACichy}. Below we present further numerical results concerning the evolution from the weak (BCS like) to the strong coupling limit of tightly bound LP's with increasing $|U|$, for $d=2$ and $r\neq 1$. The system of self-consistent equations \cite{Kujawa, Kujawa5} has been solved numerically. The first order transition lines were determined from the condition $\Omega^{SC} =\Omega^{NO}$, at fixed chemical potential, where $\Omega^{SC}$ and $\Omega^{NO}$ are the grand canonical potentials of SC and NO states, respectively. Then, these results have been mapped onto the case of fixed $n$. The diagrams have been obtained mostly for low $n$. We use $t$ as the unit. 

\begin{figure}[h!]
\includegraphics[width=0.3\textwidth,angle=270]{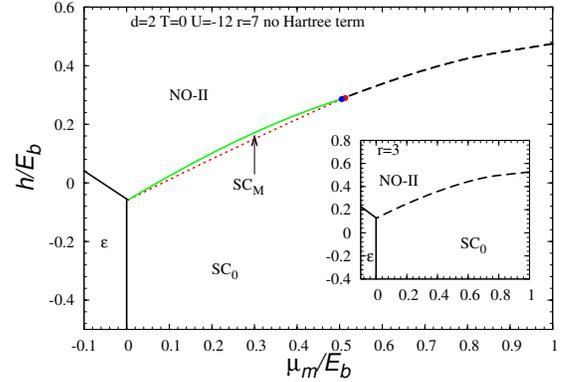}
\caption{Critical magnetic field vs. chemical potential diagram for $d=2$ at fixed $U=-12$, $r=7$ and $r=3$ (inset). $SC_0$ -- unpolarized SC state with $n_{\uparrow}=n_{\downarrow}$, $SC_M$ -- magnetized SC state, NO-II -- fully polarized normal state, $\varepsilon$ -- empty state, $\mu_m$ -- half of the pair chemical potential defined as: $\mu_m=\mu -\epsilon_0+\frac{1}{2}E_b$, where $\epsilon_0=-4t$, $E_b$ is the binding energy for two fermions in an empty lattice. Red point -- $h_{c}^{SC_M}$, blue point -- tricritical point. These points are close to each other for chosen parameter values. The dotted red and the solid green lines are continuous transition lines. The dashed black line is the 1$^{st}$ order transition line.} 
\label{fig1}
\end{figure}

\begin{figure}[h!]
\includegraphics[width=0.3\textwidth,angle=270]{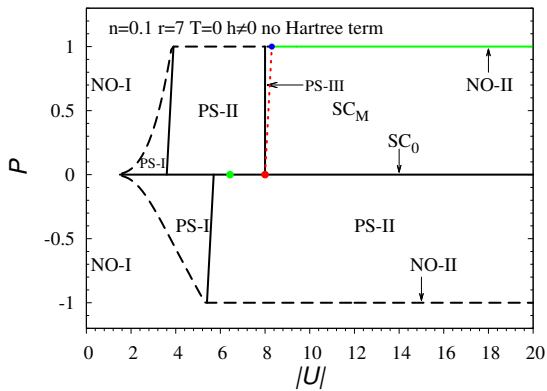}
\caption{Polarization vs. on-site attraction ground state phase diagram at fixed $n=0.1$ and $r=7$, for the square lattice. $SC_0$ -- unpolarized SC state with $n_{\uparrow}=n_{\downarrow}$, $SC_M$ -- magnetized SC state, NO-I (NO-II) -- partially (fully) polarized normal states. PS-I ($SC_0$+NO-I) -- partially polarized phase separation, PS-II ($SC_0$+NO-II) -- fully polarized phase separation, PS-III -- ($SC_M$+NO-II). Red point -- $|U|_{c}^{SC_M}$ (quantum critical point), blue point  -- tricritical point, green point -- the BCS-BEC crossover point in the SC$_0$ phase.}
\label{fig2}
\end{figure}

\begin{figure}[t!]
\includegraphics[width=0.3\textwidth,angle=270]{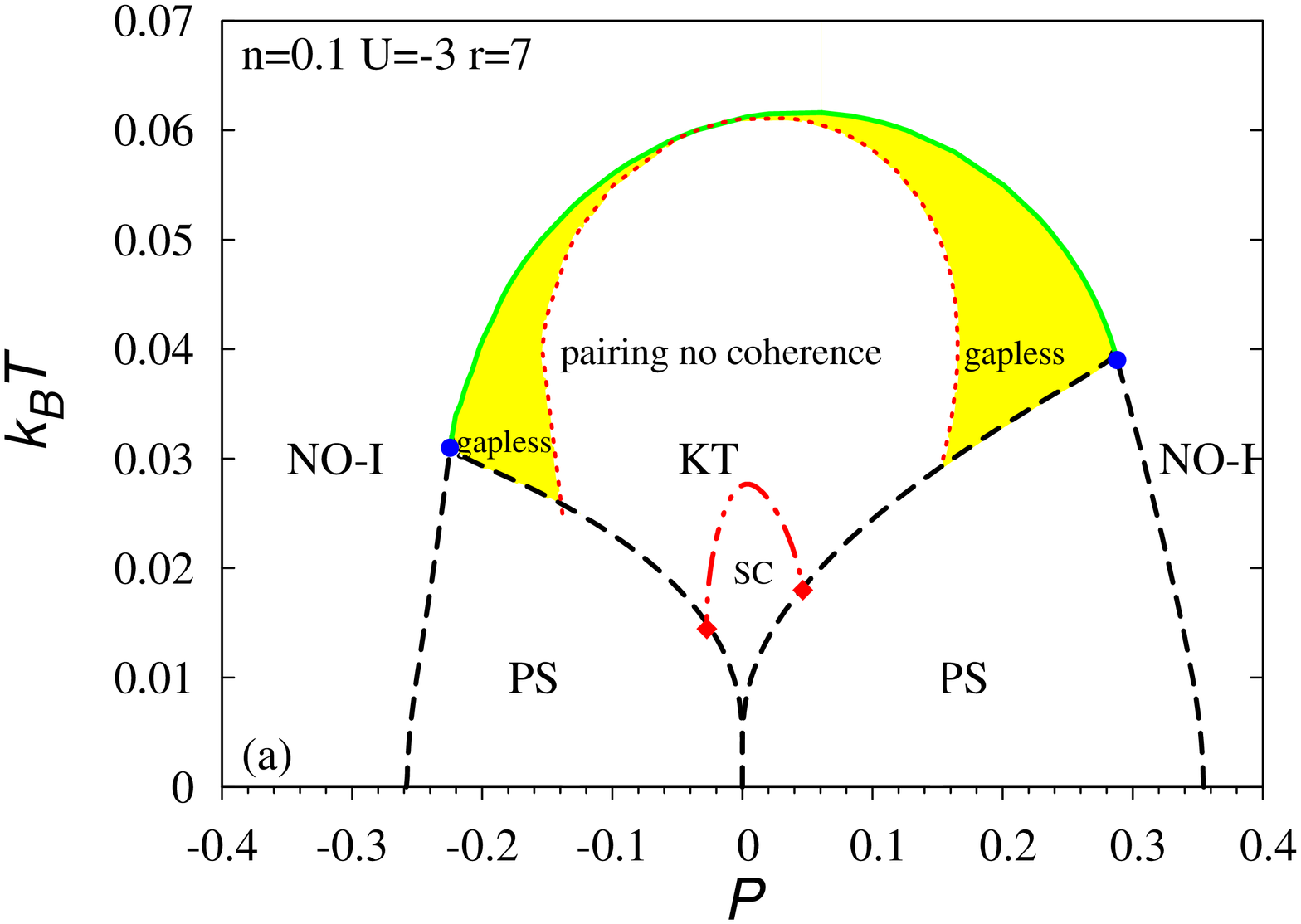}
\includegraphics[width=0.3\textwidth,angle=270]{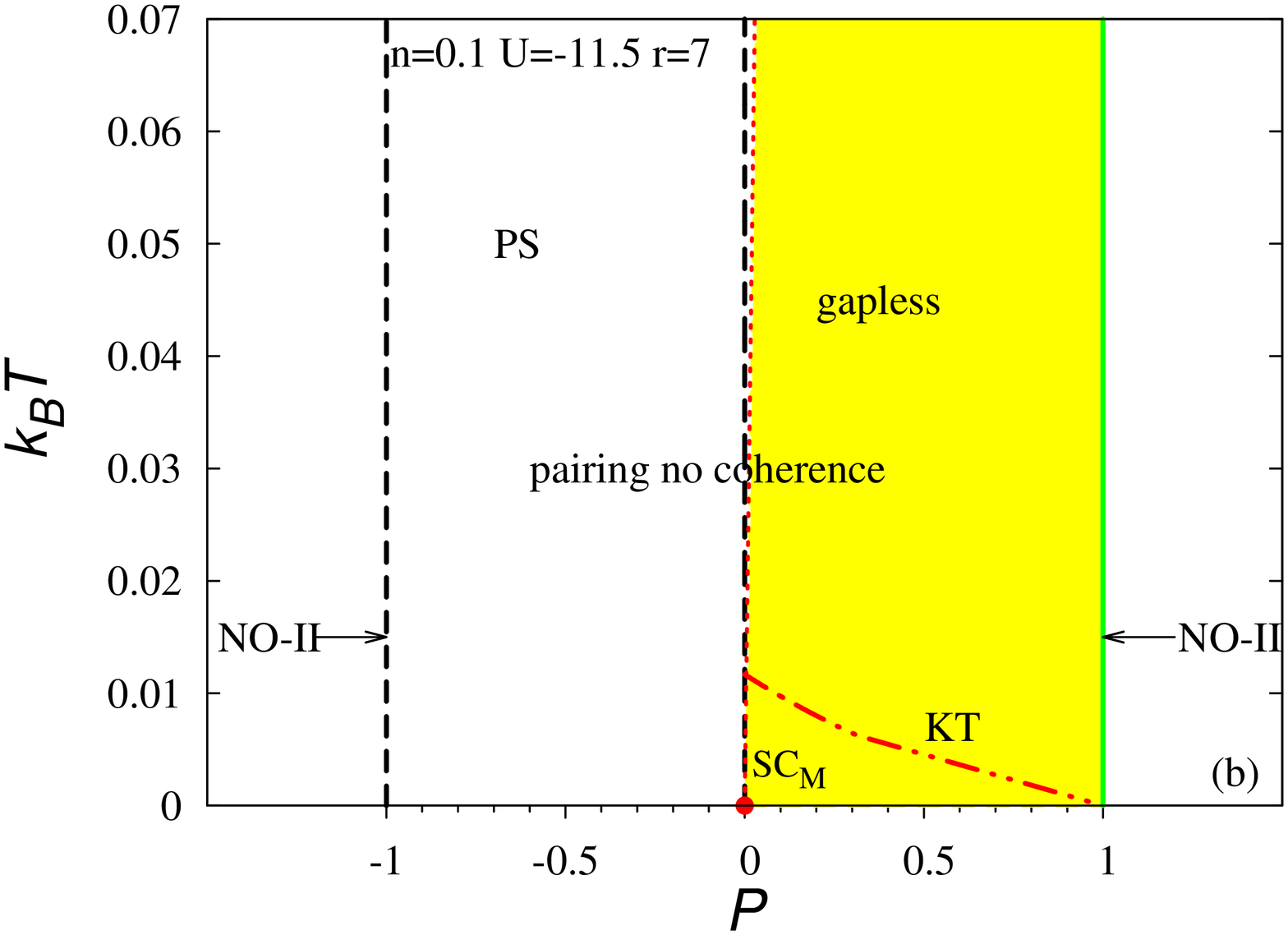}
\caption{\label{fig3} Temperature vs. polarization phase diagrams without the Hartree term, at fixed $n=0.1$, $r=7$, for the square lattice. (a) $U=-3$, (b) $U=-11.5$. The thick dashed-double dotted line in (red color) is the KT transition line. Thick solid line denotes transition from pairing without coherence region to NO within the BCS approximation. Above the dotted line (red color) -- gapless (yellow color) -- the region which has a gapless spectrum for the majority spin species. $SC$ -- 2D KT superconductor, SC$_M$ -- gapless KT SC with one FS in the presence of polarization (a spin polarized KT superfluid).}
\end{figure}

Fig. \ref{fig1} shows the phase diagrams at fixed $\mu_{m}$ and $h$, on the LP side. We define $\mu_{m}=\mu-\epsilon_0+\frac{1}{2} E_b$ as one half of the pair chemical potential (molecular potential). It is worth mentioning that in 2D system at $r=1$ the Sarma or the breached phase is unstable even in the strong coupling limit. In Fig. 1 (inset) there is only first order phase transition from pure SC$_0$ to the NO phase (with increasing $h$) at fixed $r=3$. However for higher value of the hopping ratio ($r=7$), we observe also a continuous phase transition from SC$_0$ to SC$_M$, with decreasing the chemical potential and increasing magnetic field. The character of transition from the superconducting to
the normal phase changes with decreasing $\mu$. Hence, we also find a tricritical point (TCP) in the diagram (blue point), at which a second order transition from SC$_M$ to NO-II terminates. We can have the following sequences of transitions: SC$_0$ $\rightarrow$ NO-II or SC$_0$ $\rightarrow$ SC$_M$ $\rightarrow$ NO-II. The SC$_0$ $\rightarrow$ SC$_M$ is a topological quantum phase transition (Lifshitz type). There is a cusp in the order parameter vs. magnetic field plots (for fixed $n$, for $\mu$ vs. $h$ as well). There is also a change in the electronic structure. In SC$_0$ phase there is no FS but in SC$_M$ state is one FS for excess of fermions. On can notice that the presence of the Hartree term restricts the range of occurrence of the SC$_M$ phase except for a very dilute limit.

In Fig. \ref{fig2} we present the $P-|U|$ ground state diagram for low electron concentration $n=0.1$ and fixed $r$, at $h\neq 0$. As mentioned before, in 2D system at $r=1$, for $h\neq 0$, the SC$_M$ phase is unstable even in the strong coupling limit and the phase separation (PS) is favourable. This is in opposition to the 3D case in a Zeeman magnetic field \cite{kujawa2} in which for $r=1$ the SC$_M$ phase occurs for strong attraction and in the dilute limit. However, for $r\neq 1$ SC$_M$ in $d=2$ can be stable. These types of solutions (with $\Delta (h)$) appear (for $r> 1$) when $h>(\frac{r-1}{r+1})\bar{\mu}+2\Delta \frac{\sqrt{r}}{r+1}$ (on the BCS side) or when  $h>\sqrt{(\bar{\mu}-\epsilon_0)^2+|\Delta |^2}-D\frac{r-1}{r+1}$ (on the LP side) \cite{ACichy}. In the weak coupling limit the Sarma phase (BP-2) is unstable at $T=0$ and PS is favoured for a fixed n.  However, there is a critical value of $|U_c|^{SC_M}$ (red point in the diagram), for which the SC$_M$ state becomes stable, instead of PS. The transition from SC$_M$ to NO can be accomplished in two ways for fixed $n$: through PS-III (SC$_M$+NO-II, where NO-II -- fully polarized normal state) or through the second order phase transition for higher $|U|$. The change of the character of this transition manifests itself through TCP. Therefore, the magnetized superconducting state is stable only on the BEC side, for $r\neq 1$ in $d=2$ (see Fig. \ref{fig2}, $P>0$). If $r\neq 1$, the  symmetry with respect to $h=0$ is broken. Hence, the diagram is not symmetric with respect to $P=0$ and for $P<0$ the PS is favorable instead of SC$_M$ in LP limit (Fig. \ref{fig2}). It is worth to mention that the presented phase diagram has been constructed without the Hartree term because such a term restricts SC$_M$ occurrence to a very dilute limit. 

We have also extended our analysis to the finite temperatures. 
The KT phase transition is revealed by the universal jump of the superfluid density (\ref{KT})-(\ref{ro_s}). 

Fig. \ref{fig3} shows $T-P$ phase diagrams for $n=0.1$, $r=7$, at $h\neq 0$ and two values of the attraction -- moderate weak ($U=-3$, $E_b/E_F=0.024$) and strong ($U=-11.5$) coupling. These diagrams have been constructed within the mean field approximation (the solid lines (2$^{nd}$ order transition lines), PS and gapless regions), but the phase coherence temperatures have been obtained within KT scenario (thick dash-double dotted line (red color)). In a strict theory, below KT temperature ($T^{KT}_c$) the system has a quasi-long-range (algebraic) order. In our approach this is characterized by the non-zero gap ($\Delta \neq 0$) and non-zero superfluid stiffness ($\rho_s \neq 0$). In a weak coupling regime the KT superconductor (SC) exists at low $|P|$ and low $T$ (Fig. \ref{fig3}(a)). The SC phase is restricted to low $|P|$, while for larger $|P|$ the PS region is favored. There is also the region of pairs without coherence, i.e. nonsuperfluid, formally defined by $\Delta \neq 0$, $\rho_s=0$. In this region one observes a pseudogap behavior. Therefore, the region of incoherent pairs is different from the normal phase. On the $T-P$ diagrams (Fig. \ref{fig3}(a)), one finds MF TCPs at which the thermal transition changes from the second to the first order. We also show the gapless area within the state of pairing without coherence. This gapless region is above KT coherence temperatures in the weak coupling limit.  

In the intermediate coupling, below $T^{KT}_c$ the SC is strongly reduced to very low $|P|$. In the BCS-LP crossover point a polarized SC does not exist even for $r\neq 1$. 

The situation is radically different in the spin asymmetric hopping and strong coupling case (Fig. \ref{fig3}(b)). For sufficiently high value of $r$, below $T_{c}^{KT}$ curve, a spin polarized KT superfluid state with gapless spectrum and one FS can be stable for all $P>0$. If $P<0$ there is a PS region at low $T$. 

In the strong coupling limit, $T_c^{KT}$ does not depend on the magnetic field, but it depends on the hopping (mass) ratio: $k_{B}T_c^{KT}=2\pi\frac{r}{(1+r)^2}\frac{t^2}{|U|}n(2-n)$ ($r>0$, $n<n_c$) \cite{ACichy}. This estimate for hard-core bosons (Hamiltonian (\ref{pseudospin})) gives upper bound on transition temperature in the strong coupling limit. The results obtained from Eqs. (\ref{ro_s})-(\ref{KT}) in a deep bosonic regime are in very good agreement with those obtained from the strong coupling expansion.

\section{Conclusions}

We have investigated the effect of a Zeeman magnetic field and the hopping imbalance on the BCS-BEC crossover at $T=0$ and finite temperatures, for the 2D attractive Hubbard model. 

We have obtained that if $r=1$, the SC$_M$ phase is unstable in $d=2$ even on the LP side. The effect of Zeeman magnetic field and hopping asymmetry combination (population and mass imbalance) can stabilize SC$_M$ phase on the LP side of crossover. This magnetized superfluid state, occurring for strong attraction and lower filling, is characterized by one FS and a gapless spectrum for the majority spin species. At $T=0$, the BP-2 phase is unstable both for $r=1$ and $r\neq 1$ on a 2D lattice. 

We have also extended the BCS-LP crossover analysis to finite temperatures in $d=2$ by invoking the KT scenario. In a weak coupling regime the spin polarization has a destructive influence on the KT superfluid state at $r=1$ and $r\neq 1$. However, we have found that the range of $P$ for occurrence of a spin-polarized KT superfluid is much larger on the LP side.

\begin{acknowledgments}
I would like to thank R. Micnas for guidance and many valuable discussions.
I acknowledge financial support under grant No. N N202 030540 (Ministry of Science and Higher Education -- Poland).

\end{acknowledgments}

\bibliography{cichy}

\begin{thebibliography}{23}
\bibitem[{prz()}]{przypis1}
\bibinfo{note}{In Eq. (4) of Ref. \cite{ACichy}, in the constant energy term
  $J_0$ should be replaced by $K_0$.}
\end{thebibliography}

\end{document}